\documentclass[runningheads]{svmult}

\usepackage{makeidx}   
\usepackage{graphicx}  
\usepackage{subeqnar}  
\usepackage{multicol}  
\usepackage{physprbb}  
\makeindex             

\begin{document}
\title*{{\it e-VLBI...} a Wide-field Imaging Instrument with
    Milliarcsecond Resolution \& microJy Sensitivity} 

\toctitle{{\it e-VLBI...} a Wide-field Imaging Instrument with
    milliarcsecond Resolution \& microJy Sensitivity 
}
%
%
\titlerunning{e-VLBI}
%
\author{M.A. Garrett\inst{1}}

\authorrunning{M.A. Garrett}
%
%
\institute{
Joint Institute for VLBI in Europe (JIVE), \\
Postbus 2, 7990~AA Dwingeloo\\
The Netherlands.} 

\maketitle              

\begin{abstract}
  The European VLBI Network (EVN) is in the process of establishing an
  e-VLBI array in which the radio telescopes and the EVN correlator at
  JIVE are connected in real-time, via high-speed national fibre optic
  networks and the pan-European research network, G\'{E}ANT. This paper
  reports on recent test results, including the production of the first
  real-time e-VLBI astronomical image. In a parallel and related
  development, the field-of-view of VLBI is also expanding by many
  orders of magnitude, and the first results of deep, wide-field
  surveys capable of detecting many sources simultaneously are
  summarised. The detection of sources as faint as 10 microJy should
  soon be possible in the era of ``Mk5'' and e-VLBI.
\end{abstract}

\section{Introduction to e-VLBI} 

The application of fibre optic network technology to existing radio
telescope facilities (e.g. EVLA \& e-MERLIN), will permit these
instruments to achieve sub-microJy continuum sensitivity noise levels
in very modest integration times (e.g. 12-24 hours). In addition,
new telescopes such as LOFAR will employ advanced digital processing
techniques, in order to survey huge areas of the sky simultaneously via
independently steerable, so-called ``multiple'' beams.

The VLBI community around the world is also engaged in the first
attempts to connect VLBI antennas and correlator centres in real-time
({\it e-VLBI}) using commercial optical fibre networks (e.g.
Whitney 2003, Parsley et al. 2004). In addition, improvements in data storage
media and the emergence of affordable PC clusters, are poised to
transform VLBI into a wide-field, all-sky survey instrument.  This
paper presents some recent results in these areas, describing technical
and scientific developments that are relevant to the SKA.


\begin{figure}[t]
\begin{center}
\includegraphics[width=1.0\textwidth]{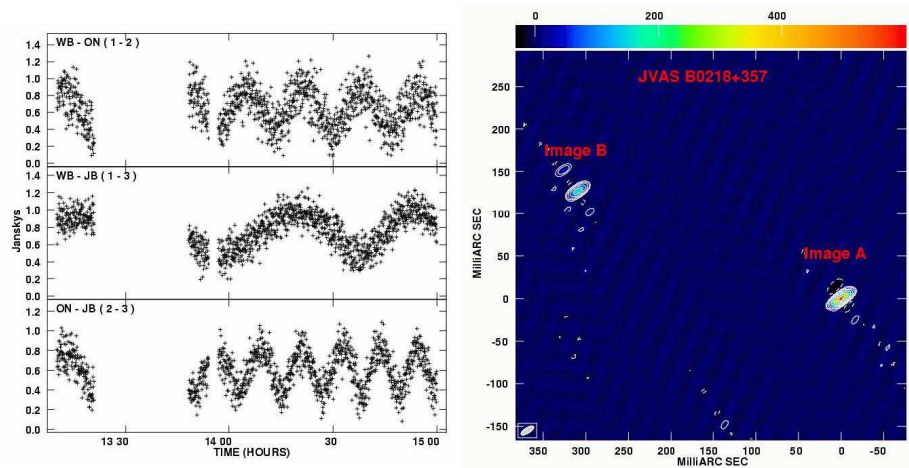}
\end{center}
\caption[]{The first real-time e-EVN image. The target source was the
  well-known, small-separation gravitational lens system, B0218+357.
  The lensed images are separated by $\sim 334$ milliarcseconds. }
\label{fig1}
\end{figure}

\section{e-VLBI: The first real-time e-EVN Image}

By definition, VLBI antennas and correlators are separated by many
hundreds, indeed thousands of kilometers. Typically they are also
located in remote areas, far from centres of population and network
services. The goal of connecting together these antennas and
correlators is therefore challenging, especially as it also requires
connections to be made across national and international boundaries.
Despite these difficulties, the prospect of achieving this goal is
becoming an increasingly realistic proposition, as advanced national,
continental and trans-continental research networks begin to
interconnect across the globe.

In Europe, the EVN has entered into a collaboration with the major
European National Research Networks (NRENs) and the pan-European
research network G\'{E}ANT, operated by DANTE. Progress has been aided by
the introduction of the new Mk5 PC-based recording systems (Whitney
2003) at the major telescopes across the EVN, including those with good
network connections. The first eVLBI tests in Europe began in September
2002, and recently culminated in observations that have included three
telescopes (Westerbork, Onsala \& Jodrell Bank) simultaneously
transmitting data at rates of 32 Mbps to the EVN correlator at JIVE.
The first real-time e-VLBI image (see Fig.~1) was produced by this
array in April 2004 (Parsley et al. 2004), the data flowing without
interruption from the telescopes to the correlator, where fringes were
generated at JIVE in real-time. After real-time correlation, the data
were automatically processed using the EVN data pipeline.  Some plots
of the visibility data and the pipelined image are presented in Fig.~1.

\begin{figure}[t]
\begin{center}
\includegraphics[width=1.0\textwidth]{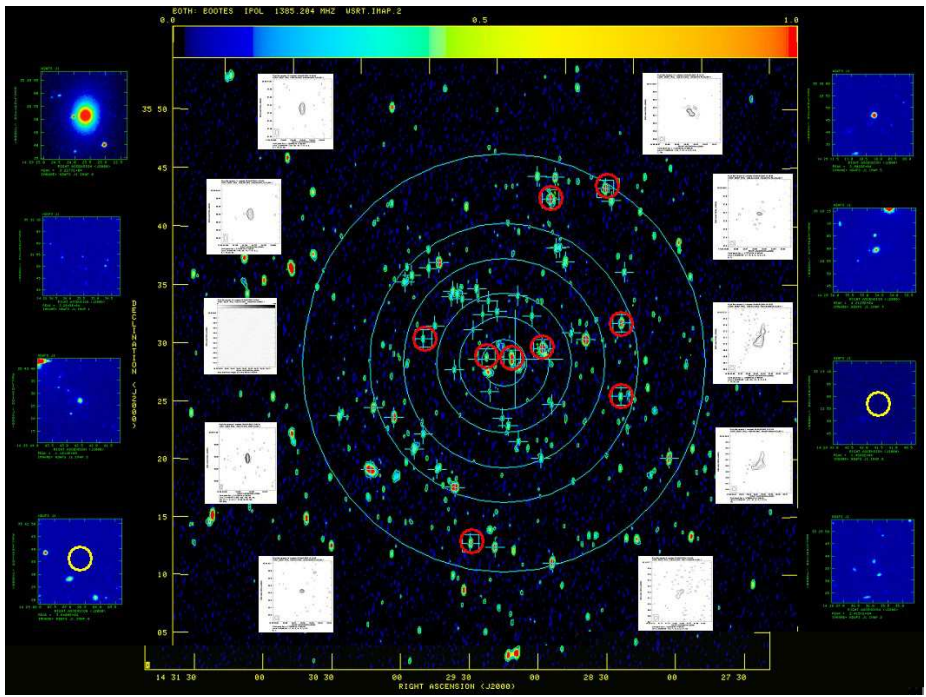}
\end{center}
\caption[]{A wide-field VLBA+GBT survey of part of the NOAO-N Deep
  Field region. With an r.m.s. noise level of 9~$\mu$Jy, many sources
  are detected simultaneously. A significant fraction of radio sources
  are not detected in the optical images also presented.}
\label{fig1}
\end{figure}

In terms of local logistics and network reliability, e-VLBI offers many
advantages over conventional VLBI data transport systems. From the
astronomers perspective, the prospect of immediate results and
unlimited access to the high data rates (currently restricted by the
available disk resources) are also important new features. Progress in this
area continues as other EVN telescopes come ``on-line'', including the
Torun and Medicina 32-m telescopes.

\section{VLBI Sensitivity and Deep Wide-Field VLBI Surveys} 

The current sensitivity of existing VLBI arrays is ``embarrassingly''
good. The new Mk5 system currently being introduced across the EVN,
permits data rates of 1 Gbps to be recorded robustly and without error.
The expectation is that a global VLBI array, also equipped with Mk5,
could achieve $1-\sigma$ r.m.s. noise levels better than a few microJy
per beam for an on-source integration time of $\sim 24$ hours. At these
sensitivity levels, VLBI can expect to simultaneously detect many
sources within the primary beam of an individual antenna (see Fig.~2).
Currently deep, wide-field observations (Garrett, Morganti \& Wrobel
2004), achieve noise levels of 9~microJy/beam and are capable of
detecting mJy, sub-mJy and microJy radio sources across the field.  The
first results indicate that the fraction of radio sources (AGN)
detected by VLBI falls from $\sim 29\%$ at mJy levels to only 8\% at
sub-mJy flux density levels. The results are in good agreement with
less direct studies -- these suggest the emergence of a dominant
star-forming radio source population at these faint flux density
levels.

Fig.~3 shows an example of the compact radio sources that might be
simultaneously detected in a typical region of sky by a Mk5 equipped
Global VLBI array, assuming the fractional detection rate of 8\% (as
observed for sub-mJy sources) is also appropriate for the microJy radio
source population. Current and future wide-field VLBI surveys are
likely to be highly efficient AGN detectors, and in addition, may be
sensitive to cold, very high-redshift, dust-obscured systems that will
be difficult to detect in other wave-bands. 

\begin{figure}[t]
\begin{center}
\includegraphics[width=1.0\textwidth]{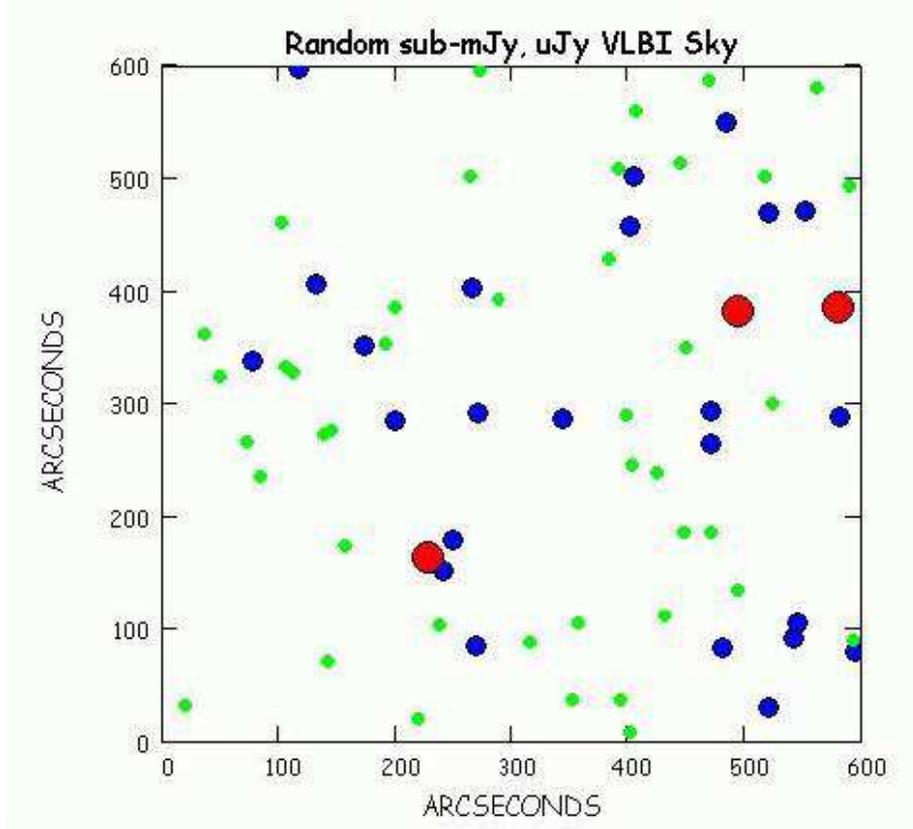}
\end{center}
\caption[]{A view of the faint but compact (AGN) radio sky as viewed by
  a wide-field global VLBI array. Representation: large dots ($S >
  1$~mJy), medium dots ($S \sim 100-1000 \mu$Jy) and small dots ($ S
  \sim 10-100 \mu$Jy).}
\label{fig1}
\end{figure}

%

\end{document}